# Carbon Nanotubes Production Using Arc Ignition Under Magnetic Field

V.K. Jindal[1], K. Dharamvir, Vitisha Suman and Mepam Tsomo
Department of Physics, Panjab University, Chandigarh-160014, India
email: jindal@pu.ac.in

We demonstrate that AC arc discharge under de-ionized water which has been recently used for carbon nanotubes (CNT) production can be modified to increase the quantity and quality of carbon nanotubes by applying a magnetic field. The carbon soot thus obtained was analyzed using SEM, TEM and Raman spectroscopy. The Raman scattering measurements showed interesting results. Radial breathing modes (RBM) were observed in the Raman data in the range 150 - 700 $cm^{-1}$. The presence of RBM is good enough indication of the presence of the carbon nanotubes in the soot collected by us. The RBM data was analyzed further which indicated a superposition of two closely lying modes. It showed the distribution of tube diameter which depended on the application of magnetic field.

Iijima [1] had discovered the nanoscopic threads lying in the smear of soot, produced by the arc discharge method in early 1990s. These nanoscopic threads were made of pure carbon, had regular and symmetric structures as crystals, and were later named Carbon Nanotubes (CNTs). Since then, the carbon nanotubes are continuously in the focus of attention as they hold significant promises for a vast number of material applications due to their unique mechanical, electrical and storage properties.

The production of carbon nanotubes is an important area of activity as there is an increased demand of this material for various applications and research. Even after more than a decade of its discovery we still do not have a method to produce carbon nanotubes of required characteristics. Arc method was the first method used for production of macroscopic amounts of carbon nanotubes [2]. Since then new methods have been designed and the existing ones were modified [3-5]. Recently production of carbon nanotubes under inert liquid mediums specially de-ionized water and liquid nitrogen has engrossed a lot of attention in the scientific community [6-8]. Lately a method for CNT production was reported by Biro et al. [9] using an underwater AC arc ignition of graphite. This method employs the benefits of an AC and the arc is carried out under de-ionized water leading to successful production of carbon nanotubes.

---

[1] Corresponding author

In this paper we report the production of carbon nanotubes working on the idea similar to that of Biro et.al. We have suggested some modification, mainly by igniting the arc under a magnetic field. The motivation for using magnetic field is that it provides the initial Lorentz force to the ionized carbon atoms which are produced during arc ignition of graphite, thus probably providing possibility of giving a curvature leading to enhancing tubular formation in preference to parallel sheets.

In the proceeding sections first the experimental details are given, followed by the analysis carried out using scanning electron microscopy (SEM), transmission electron microscopy (TEM) and Raman spectroscopy. Then results are followed by some discussion and conclusion.

**Experimental Details**

An arc discharge typically uses a chamber filled with an inert gas at low pressures. To execute the arc between say, graphite rods, traditionally a DC voltage around 20-30 V is applied designed to draw currents around 100-200 Amps. In the present case, the arc excitation between graphite rods is carried out under de-ionized water instead of inert gas. Further, AC power supply is used to provide 100-200 Amps at 20-50 V.

Figure 1 shows the schematic drawing of the arc discharge apparatus used in our experiment. AC power supply was set at 50V and 200A. The anode and cathode rods used were pure carbon rods of 6 mm diameter each. The trough made of thick glass was filled with de-ionized water; so that the graphite arc is produced underneath water. The idea of using de-ionized water follows the work first reported by Biro et al [8]; this seems to reduce the amount of debris in the soot. The production method was repeated by us by applying varying values of magnetic field perpendicular to the graphite electrodes using electromagnets having a soft iron core and copper wire windings around the core. Each pole had 16 mm diameter and 1200 turns, insulation was provided between each layer to prevent short circuiting and damage.

.

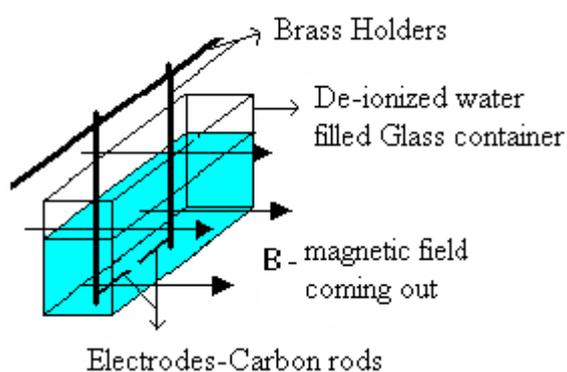

**Figure1.** Schematic experimental set-up, Electromagnet (not shown) is used to apply magnetic field.

The electrodes were horizontally aligned and during arc ignition, their distance was manually controlled so that they always remain at a separation of approximately 1 mm to obtain a continuous and beautiful arc with sputtering of black colored soot in all direction below water. The movement of the carbon rods is managed using a gap control screw unit. When a considerable amount of carbon rods were consumed and arc discharge was over, the carbon soot was separated from the water and was dried. Then it was taken for the further analysis.

**Characterization and Results**

### A. Scanning Electron Microscopy (SEM)

Scanning electron microscopy was done to study the constituents of the carbon soot. The specimen was dried in the desiccator and was then coated with Au prior to SEM observation. Fibrous structures as were observed in the SEM image are shown in the Figure 2. The SEM pictures indicate that we do have some long rod like structures. The scale in the Figure2 (a)-(d) is indicative of the thickness of the rod like structure is around 25nm ± 5nm and length is of order of few micrometers. Further form the pictures it can be seen clearly that with application of magnetic field the production of the rod like structure has increased and the aspect ratio gets enhanced under magnetic field. Here the magnetic field plays a crucial role in enhancing the yield.

### B. Transmission Electron Microscopy (TEM)

We carried out TEM analysis of the samples using TEM available at Institute of Physics, Bhubaneshwar. Samples for Transmission Electron Microscopy were deposited onto 300 mesh copper TEM grids coated with 50nm carbon films. After evaporation by sonication, the grids were examined in JEOL 2010 microscope with Ultra-High Resolution microscope operated at 200 kV. HRTEM images were obtained. Some of these are presented in Figure. 3. Unfortunately, the TEM analysis was carried out only for the samples obtained in first batch of production when no magnetic field was applied. We can notice that several multiwall carbon nanotubes are identified in the figures.

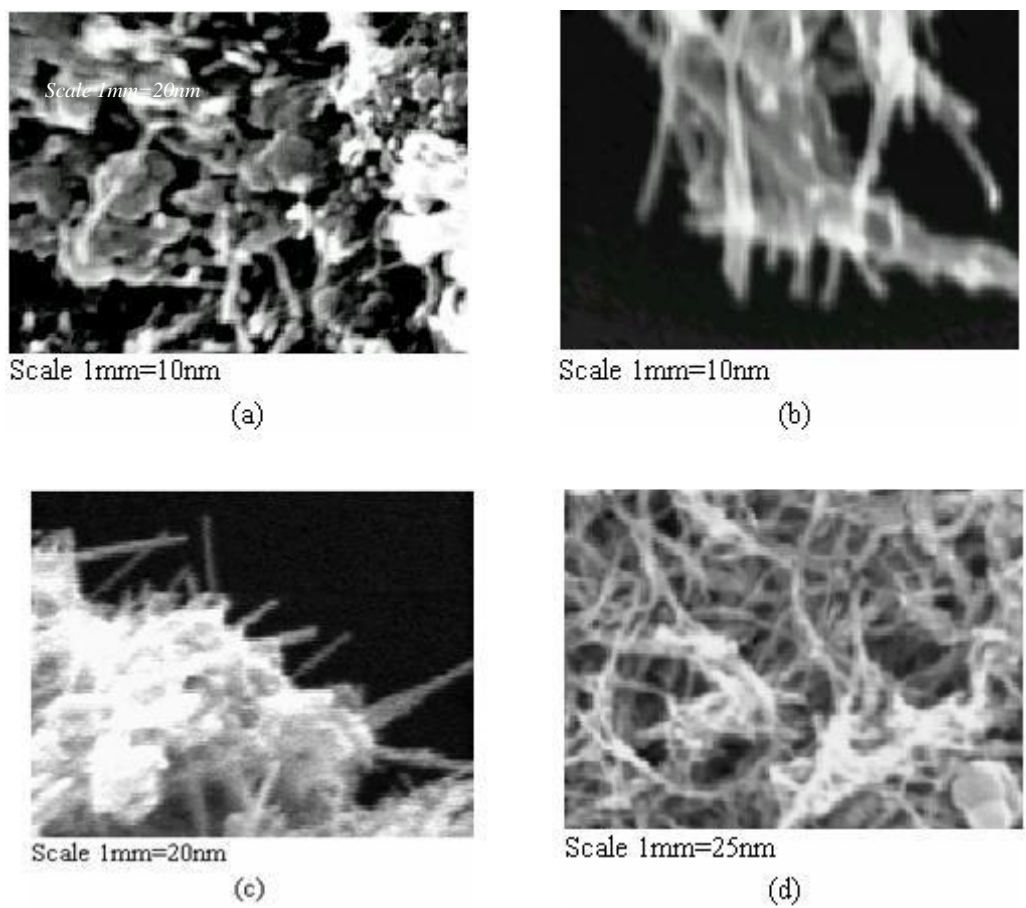

**Figure 2.** SEM images of MWNTs synthesized by arc discharge method in de-ionized water (a) without magnetic field (b) with magnetic field **B** = 0.023 T. (c) and (d) with magnetic field **B** = 0.026 T.

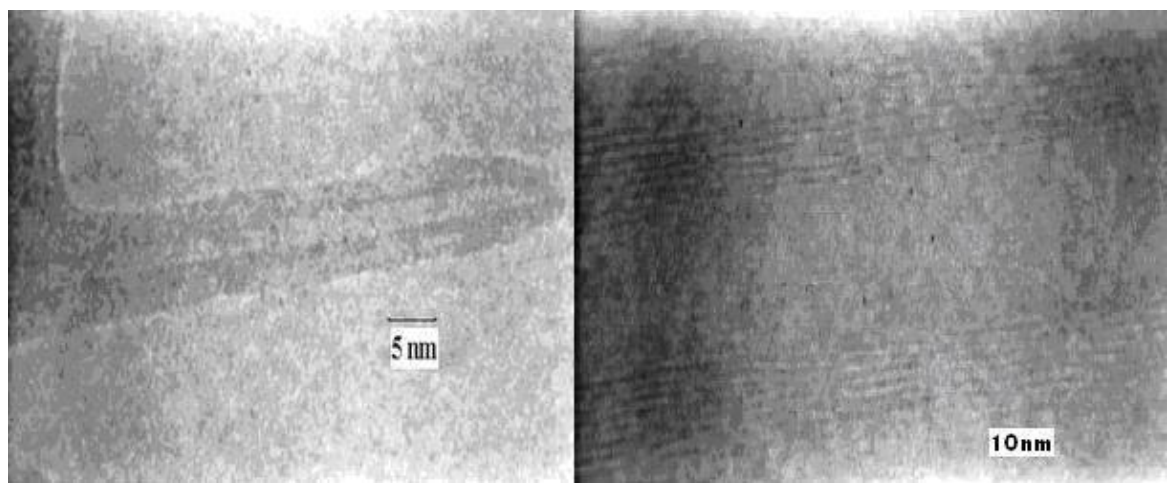

**Figure 3.** TEM photographs showing multiwall nanotubes (B=0).

### B. Raman Measurements

We measured the Raman spectra of the soot collected using the facilities LABRAM, using laser of 514.532 nm and 50 mw at University of Wuerzburg, Germany. In Raman spectra the three regions of interest are (i) Radial Breathing Mode (RBM), this is the frequency range below 800 cm$^{-1}$, (ii) D-mode around 1330 cm$^{-1}$ and (iii) G-mode close to 1580 cm$^{-1}$. The spectra have been presented in Fig. 4. We now discuss these spectra.

**(i) Radial Breathing Mode measurements**

The most important feature in the Raman spectra of CNTs is the Radial Breathing mode. RBM in Raman spectra is unique to cylindrical symmetry. The frequency of the RBM is directly linked with the tube diameter. This fact has been widely used to calculate the diameter of SWNTs [11-12].

RBM of MWNTs has been reported by Y. Ando et al. [13]. It was later confirmed that the RBMs of MWNTs originate from the very thin innermost tube [14]. In this paper we also report the distribution of innermost diameter we have calculated. For isolated SWNT diameter is inversely proportional to the RBM frequency [12],

$$d = 223.75/\omega \qquad (1)$$

where, d is the diameter of the isolated SWNT in nanometers and ω is the RBM frequency in wave numbers. No RBM peaks were seen for the carbon nanotubes produced in the absence of magnetic field. For the samples produced in the presence of magnetic field, the RBM peaks are clearly visible but somewhat smeared out.

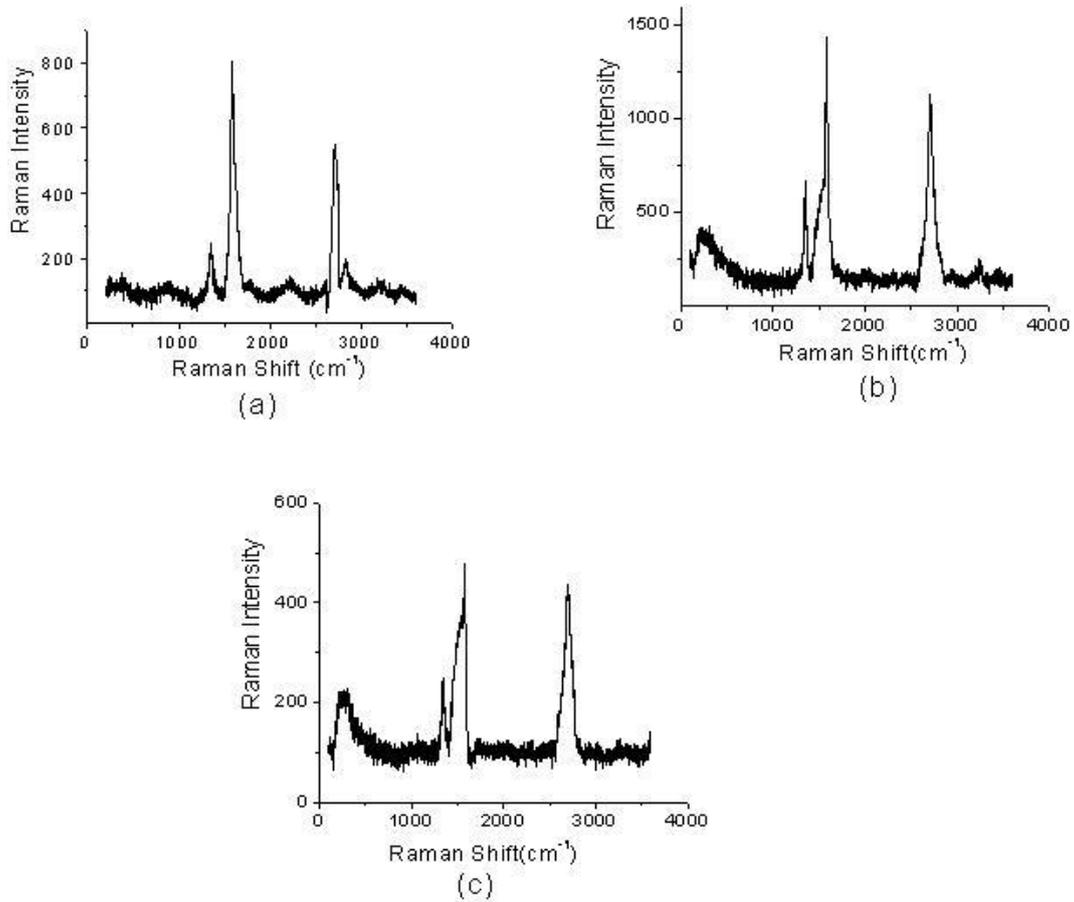

**Figure 3.** Raman spectra of three MWNTs samples (a) without magnetic field, (b) with magnetic field **B** = 0.023 T, (c) with magnetic field **B** = 0.026 T.

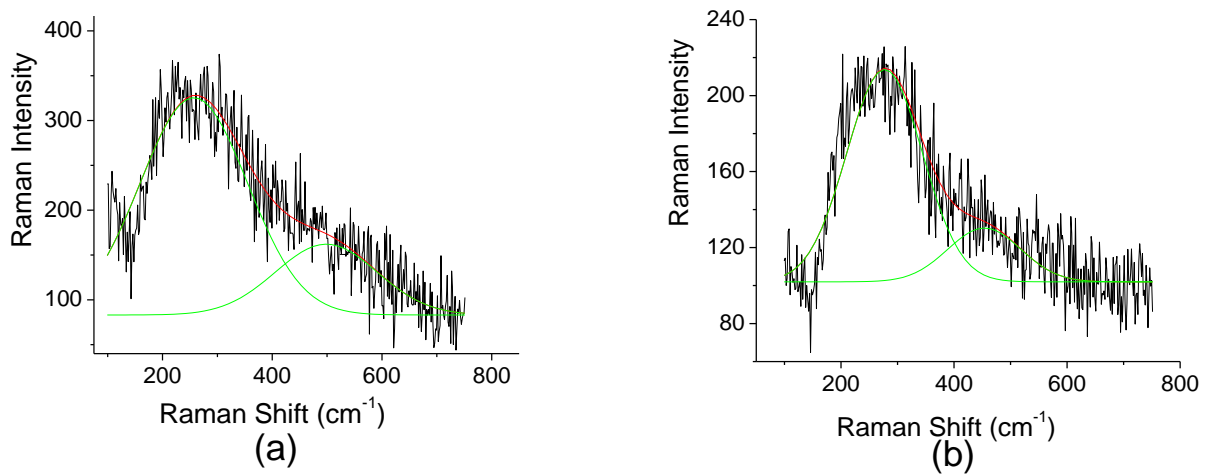

**Figure 4.** Raman spectra RBM region (a) with magnetic field **B** = 0.023 T, (b) with magnetic field **B** = 0.026 T. Both spectra are fitted with Gaussian functions that are attributed to dominant and non-dominant peaks respectively.

The presence of the RBM peaks gives a confirmation for the presence of some cylindrical entities in the soot which we attribute to the presence of carbon nanotubes in our produce. Moreover since the RBM peak is not sharp and is smeared out it indicates that the amount of MWNTs is dominating in the soot of SWNTs. For the samples which were synthesized under two slightly different values of the magnetic field, the RBM peaks appear as if these are a superposition of double broad peaks. This can be observed from Fig. 4, where the RBM peaks are focused separately. We also did the Lorentzian fitting to these peaks and are also shown in the same Fig.. Form the Lorentzian fitting it becomes clear that the RBM peak is superposition of two broad peaks; we have been able to fit the composite RBM peak by different percentage of the constituent peaks. The comparison for the diameter distribution of the CNTs in the samples with magnetic field **B** = 0.023T and **B** = 0.026T is given in the Table1.

**Table1**. Analysis of the diameter of carbon nanotubes based upon RBM. Here, d diameter of the tube, ω is the position of the RBM peak as picked from Figure. 4, Δd is the diameter width.

| Magnetic field, **B** (Tesla) | Dominant Peak | | | | Non- Dominant peak | | | |
|---|---|---|---|---|---|---|---|---|
| | Abundance ( %) | $d_1$ (nm) | ω (cm$^{-1}$) | Δd (nm) | Abundance ( %) | $d_2$ (nm) | ω (cm$^{-1}$) | Δd (nm) |
| 0.023 | 67 | 0.873 | 256 | 0.991 | 33 | 0.446 | 501 | 0.211 |
| 0.026 | 62 | 0.809 | 276 | 0.502 | 38 | 0.494 | 453 | 0.172 |

It can be seen from the Table 1 that our peaks in RBM for lower magnetic field are distributed around CNTs of diameter $d_1$ around 0.873 nm and $d_2$ around 0.446 nm with abundance of 67% and 33% respectively. For higher magnetic field value this distribution is around 0.809 nm and 0.494 nm with abundance 62% and 38% respectively. The tubes in both samples are mainly distributed over two ranges represented by the dominant and the non-dominant peaks, and the range over which the diameter varies gets reduced by increasing the magnitude of the magnetic field. An increase in magnetic field by about 15% refines the diameter distribution to narrow it down by about 1/5. Thus the quality of CNT production has a significant dependence on magnetic field. We also notice some shift in diameter value because of increased magnetic field. So presence of magnetic field can be used for optimized production of CNTs.

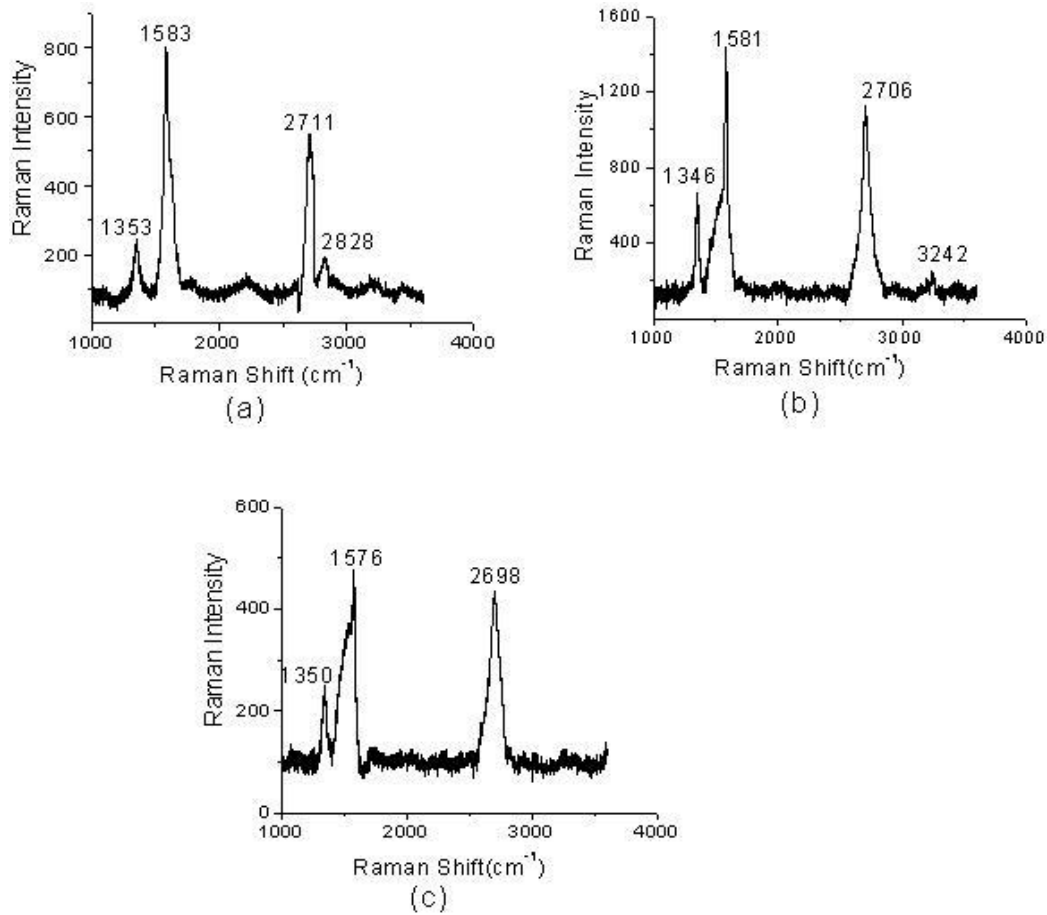

**Figure 5.** Tangential mode and D-mode (a) without magnetic field, (b) with magnetic field **B** = 0.023 T, (c) with magnetic field **B** = 0.026 T.

**(ii) G-mode measurements**

The regions of interest in Raman spectra other than the RBM mode are G-mode, the tangential mode, and D-mode (disorder mode). These have been presented in Fig. 5. The G-mode appears around 1580 cm$^{-1}$, it can also be used for diameter characterization but the information provided is less accurate [15], therefore G-mode analysis is generally not done. We are mentioning it only for the sake of completeness. The small intensity peaks around 1580 cm$^{-1}$ can be attributed to a small component of a mixture of CNT with different diameter.

D-mode, the disorder band, appears around 1330 cm$^{-1}$, it appears mainly due to defects in curved graphene sheet.[16-17].

**Discussion and Concluding Remarks**

We describe a successful production technique for CNTs as confirmed by the SEM and TEM images and the successive Raman analysis. The procedure used here is a low cost, easily up-scalable and a simple method for MWNT production. The yield of MWNTs increases with the increasing magnetic field and the Raman spectroscopic analysis shows that the variation in diameter of tubes gets reduced effectively by use of magnetic field. The spread in diameter reduces to about 20% from its spread at B=0.023 T. This remarkable increase in quality is caused by only about 15% increase in magnetic field. Thus magnetic field plays a very important role in improving the quality, i.e., providing us a way to control the wide range over which we can produce CNTs, and also in exercising a control over the required diameter. This restriction of the range over which the diameters vary in the soot seems to result in MWNTs being converted to SWNTs or that the number of the walls of the MWNTs i.e. the layering of the MWNTs is being reduced by the increase in the magnetic field. But these interpretations need further detailed analysis. We require carrying out further characterization using the available techniques like TEM. There is large scope for more work to be done by varying the magnitude of the magnetic field over a much wider range. Further, trials of production using arc-ignition of graphite in liquid nitrogen instead of de-ionized water also seem promising and will be published separately after more results are available.

**Acknowledgements**

**VKJ wishes to acknowledge research grant from Terminal Ballistics Research Lab, DRDO for carrying out this work.**